\documentclass{article}
\usepackage{graphicx} % Required for inserting images

\title{Why does the wavefunction `collapse' in relational approaches to quantum mechanics?}
 \author{Emily Adlam \thanks{Philosophy Department and Institute for Quantum Studies, Chapman University, Orange, CA92866, USA \texttt{eadlam90@gmail.com} }}
\date{\today}
\usepackage{natbib}

\begin{document}

\maketitle

\begin{abstract}

I argue that there is a straightforward way to understand the occurrence of wavefunction collapses or `quantum events' in relational approaches to quantum mechanics: we necessarily encounter a  discontinuity in our description when a system interacts with the reference relative to which we are describing it,  since the reference system cannot be described relative to itself. This makes it clear how recent concerns around quantum events in relational quantum mechanics should be resolved. However, the solution requires accepting that quantum mechanics is not a complete description of all physical facts, and moreover I argue that this is most likely inevitable if we want to be able to give a precise description of quantum events.

\end{abstract}

\section{Introduction}

Relational approaches to quantum mechanics combine two main ideas. First, that the  quantum formalism is a relational description characterizing one system relative to another. And second, that a `quantum event'  occurs when two systems interact.

This second idea is the source of many of the problems faced by relational approaches. Roughly speaking a quantum event looks rather like a wavefunction collapse and plays a role similar to measurement, but many recent critics of relational quantum mechanics have pointed out that it seems difficult to define precisely when, how and under what circumstances such a quantum event takes place  \citep{Mucino2022-MUCARQ,Calosi2024-CALRQM,Faglia-2025,Ladyman-2026}. More generally there is a worry that quantum events  do not  follow naturally from the basic commitments of the relational approach, and hence they look like a rather ad hoc move to save the phenomena. 

But in this article, I will argue that there is actually a very natural way to understand why something like a quantum event or a wavefunction collapse is inevitable in a relational approach:   if we are describing one system relative to another, we will inevitably encounter some kind of discontinuity in our description when the interaction between the system and reference becomes significant, since the reference system cannot be represented in the description relative to itself.  

Using this idea, I suggest that despite the terminology referring to `events' and `collapses,' and contra common presentations of relational quantum mechanics, quantum events should not in fact be understood as  discrete instantaneous events - rather a `quantum event' simply refers to a period of time in which it is not possible to describe the evolution of the system relative to the reference system because the system and reference are interacting strongly.  Consequently there is no need to associate a quantum event with some specific time or fixed temporal interval, because the relevant interactions are continuous and pervasive and thus it is simply a matter of deciding when the interactions have become sufficiently strong that the relational description is no longer adequate.  

The criticisms are, however, correct in one way. Relational quantum mechanics has typically been formulated so as to include a postulate that quantum mechanics is `complete,' i.e. there are no physical facts about reality outside of what can be captured in the quantum formalism. By contrast, I argue here that an adequate understanding of quantum events leads naturally to the view that the relational description provided by quantum mechanics  should be understood as  an approximation which works only as long as the system-reference interaction is sufficiently weak.  Thus I argue that once we start trying to define  quantum events in a precise way we will probably find that we must give up the idea that quantum mechanics is complete. 

Viewed in this light,  recent concerns around the definition of quantum events are not an \emph{objection} to  relational approaches so much as they are a clue as to the direction in which these approaches ought to  be developed. The idea that the quantum formalism is inherently relational is powerful and interesting in its own right, and although this idea has often been coupled with the view that quantum mechanics is complete, there is no reason it must be tied  forevermore to that position: as we will see throughout this article, the approach may become more coherent and explanatorily powerful if we allow that there may exist some absolute description which lies beyond the quantum formalism. 

I begin in section \ref{why} by discussing relational approaches to quantum mechanics in general and   explaining why something like a `wavefunction collapse' is inevitable in such an approach. Then in section \ref{RQM} I introduce relational quantum mechanics (RQM) and two variants of it, RRQM and ARQM, as well as a third variant ARQM* which explicitly gives up the assumption of the completeness of quantum mechanics. I explain how the ideas of section \ref{why} can be used to resolve concerns around the definition of quantum events in the context of ARQM*. In section  \ref{complete} I argue that it follows in a straightforward way from the basic principles of RQM that  quantum events probably cannot be derived in a precise way entirely within the standard quantum formalism, so something like ARQM* is probably the most viable way forward. Finally in section  \ref{criticisms} I respond  to some points made by \cite{Faglia-2025} and \cite{Ladyman-2026}.

\section{Why `Collapse'? \label{why}}

Before getting too deep into RQM specifically, let us begin by making some general comments about relational approaches to quantum mechanics, in order to understand why it is natural to expect something like `quantum events' or `wavefunction collapses' to occur within a relational formalism.

The central motivating idea for relational views of quantum mechanics is that quantum descriptions are relativized, and it is generally agreed that this includes at least the quantum states. But what does it mean for a quantum state to be relativized? There are a variety of approaches one might take to the metaphysics of this question,  but rather than presupposing any metaphysics, let us simply begin with a minimal, functional notion of relativization: 

\begin{quote}

The quantum state  of system $S$ relative to system $R$ is the correct state to predict the outcomes of measurements performed by $R$ on $S$, but not necessarily the correct state to predict the outcomes of measurements performed by some other system $R'$.

\end{quote}

 Here the word `measurements,' may be used in a general way to refer to some broad class of measurement-like interactions, e.g. those including quantum events. I shall leave the class unspecified for now, since the purpose of this article is precisely to understand what that class might be. 

One natural corollary of this definition is that systems probably should not be assigned quantum states relative to themselves - such states would be lacking their central defining content as codified  above, since a system cannot interact with or measure itself\footnote{Of course one degree of freedom of a system may interact with a different degree of freedom of that same system, but we will never have a degree of freedom interacting with itself.}. Note also that when we use relational descriptions elsewhere in physics  it is commonly the case that giving a description of one system relative to another amounts to stipulating that the reference system is no longer a dynamical degree of freedom \citep{article3,2021trinity} - for example, if I describe the position of a billiard ball relative to the table, that typically means I am placing the origin of my coordinate system at some fixed point of the table and thus not allowing the position of the table to vary. So if we imagine that quantum relativization works in a similar way, a system should not appear as a dynamical degree of freedom relative to itself and therefore it should not be assigned any nontrivial state within the quantum description relativized to itself. 

In addition, it can be shown that assigning systems quantum states relative to themselves leads to various paradoxes \citep{kastner2024conventionalquantumtheorydoes}, which can easily be avoided by simply not permitting systems to be assigned states relative to themselves. More generally, a whole host of philosophical complications arise when we allow self-reference \citep{IsmaelForthcoming-ISMTOU}. Thus I will henceforth suppose that it is  a natural corollary of the relational approach that quantum systems should not be assigned states relative to themselves\footnote{Of course there are alternative ways of arriving at relational descriptions which do something more complex than simply removing the reference system from the description \citep{2021trinity,rovelli2022philosophical}, so I do not mean to claim that it is impossible in principle to write down any relational description which allows for an interaction between the system and the reference. However, these more complex approaches do not typically proceed by simply writing down a state of one system relative to the other and enacting time evolution, so quantum mechanics understood as a relational description looks more askin to the approach which simply removes the reference system from the description. Indeed, it is often pointed out that the reason we have a measurement problem is precisely because standard quantum mechanics treats `observers' as external to the theory \citep{Bell1973-BELSAO}, and thus by construction quantum mechanics appears to be the kind of relational representation which leaves out the reference system rather than modelling it.}.

Suppose then that we are looking at a description of $S$ relative to $R$; what will happen if $S$ interacts with $R$? Well, since $R$ is not assigned a quantum state relative to itself, evidently the interaction between $S$ and $R$ cannot be described relative to $R$, since  $S$ is interacting with something which simply does not exist within this description. So when $S$ interacts with $R$   what we will see, within the description relative to $R$, is some kind of `singularity'  - in effect we must put down the description relative to $R$ when $S$ and $R$ begin interacting and then pick it up again once they cease to interact. Therefore within the description relativized to $R$ the effect of the interaction will be  a discontinuous, non-unitary change in the state of $S$ relative to $R$, i.e. something that looks like a `wavefunction collapse' or a `quantum event.' In this article I will refer to such a discontinuous change as an `$S-R$ quantum event' in order to connect to the existing RQM literature, although I emphasize that this term is not ideal given that, as we will shortly see, these `events' are not discrete or instantaneous. 

This is a hitherto underappreciated but very appealing feature of relational approaches in general: they have a  natural explanation for the existence of a phenomenon like `wavefunction collapse,' because they can understand it as simply the kind of model breakdown that arises from  self-reference. In particular, it is clear that such a thing  will occur during measurements. This is because it follows from  the functional definition of the relative state given above that the state used by an observer to  describe the system she is about to measure must be the state of that system \emph{relative to her} (and/or her measuring device),  otherwise the state will not give the right predictions for the outcome of her measurement. So any  measurement necessarily involves a system interacting with the reference relative to which its state is being described, giving rise to a discontinuity. Thus relational approaches allow us to understand in straightforward physical terms what is so special about `measurement': anything that we would naturally call a measurement necessarily involves a system interacting strongly with the very reference system relative to which we are presently describing it, and it is necessarily the case that something like a `wavefunction collapse' or at least a partial collapse will occur under these circumstances.

\subsection{The continuity of interaction}

At this juncture one might naturally object that interaction is not a discrete matter - for any two systems $S$ and $R$, it is plausible that $S$ and $R$ will always be interacting to some degree. For example,  in any given period of time there will necessarily be some kind of gravitational interaction between them which cannot be blocked or shielded. So does it not then follow from the account suggested above that $S$ can \emph{never} be described relative to $R$?

Yes and no. The pervasiveness of interaction does indeed mean that we probably cannot ever give a completely precise description of  $S$ relative to $R$, because there will always be some kind of interaction with $R$ itself which cannot be accounted for in the description relative to $R$. But if the interaction  is very weak then its impact may be minimal, so in many cases describing $S$ relative to $R$ may still provide a very good approximation, since  the effects of the interaction between $S$ and $R$ can be disregarded. That is, the description of  $S$ relative to $R$ encoded in the quantum formalism should be understood as a convenient approximation which works well as long as $S$ and $R$ are interacting only very weakly; when the interaction becomes stronger the approximation fails.

In order to show that this picture is not completely far-fetched, note that similar phenomena already occur elsewhere in physics. In particular, the usual interpretation of general relativity suggests that variables written as functions of coordinates cannot be physically real on their own; the physically real `absolute' quantities must be invariant under diffeomorphisms. As explained by \cite{rovelli2022philosophical}, there are several ways of getting a handle on the physical content of the theory, but one useful approach involves `interpret(ing) the coordinates as labels of concrete reference objects but ... disregard(ing) the dynamical laws governing these reference objects. The under-determination in the evolution equations can then be interpreted as the result of disregarding these dynamical laws, namely choosing physical reference systems that move arbitrarily in spacetime.' In this description, since we do not give any dynamics to the  reference relative to which we are describing a given system, it is necessarily the case that we cannot describe an interaction of the reference and the system. So we get exactly what I have described here - an approximation which remains valid as long as we do not try to have the reference objects interact with the systems under description. 

\subsection{What kind of collapse? \label{nature}}

Of course, what the argument above delivers is only the conclusion that there should exist some kind of  discontinuity in our relative description when the system interacts with the reference. By itself the argument does not explain  why the discontinuities should take some particular form - e.g. why do strong interactions have the effect of `collapsing' the wavefunction rather than effecting some other kind of non-unitary transformation, and why does a given interaction produce a collapse in some particular basis rather than another? 

Given the interpretation I have described above, it is probably the case that a complete answer to this question would require saying more about the nature of the underlying  picture to which quantum mechanics is a relational approximation.  For example, perhaps the underlying picture is formulated in terms of some  absolute quantities similar to `complete observables' in general relativity \citep{Rovelli_2002}, and a sufficiently strong interaction produces a well-defined value for a complete observable composed of a relation between $S$ and $R$, so the result of the measurement is simply a relational representation of that complete observable. 
 
But in any case, even without specifying the underlying picture we can draw some conclusions about what kinds of discontinuities we should expect to see within the relational descriptions. In particular, given the foregoing observations about the continuity of interactions we should expect to find a continuous spectrum of possible discontinuities ranging from no discernible effect when the interaction is very weak to a full collapse when the interaction is very strong. In intermediate cases, within the description relative to $R$   we have a system  undergoing unitary evolution which is, in effect, subject to a perturbation coming in from the outside that cannot be explicitly modeled. This would naturally be expected to lead to some kind of  `noisy' unitary evolution - in the description relative to  $R$ the system will undergo a non-unitary transformation, but not a full collapse. The unitary description is not completely useless and may still offer a reasonable approximation, but it becomes less and less accurate as the effects of the interaction become stronger. 

Thus interactions between system and reference which are sufficiently weak  should be expected to lead to some kind of partial collapse. And indeed, this is exactly what we do in fact see. Recently there have been significant developments  in the study of `weak measurements,' in which we do not measure a system directly but rather we entangle it with an ancillary degree of freedom and measure the ancilla instead. If we make the coupling between the ancilla and the system quite weak, then we will obtain only a small amount of information, but also the original system is disturbed only weakly \citep{jordan2024quantum}:   `the collapse is now understood to be partial, with correspondingly partial information drawn from the measurement yielding a non-unitary transformation of the quantum state' \citep{Katz_2008}. This phenomenon is theoretically well-understood \citep{kocsis2011observing,danan2013asking,jordan2014technical} and  has often been experimentally demonstrated \citep{hosten2008observation,viza2015experimentally,piacentini2016experiment}.

The phenomenon of partial collapse is very much what we would expect to find if `quantum events' arise under circumstances where the approximation encoded in the relational quantum description breaks down. In this picture, a weak measurement involves a system $S$ described relative to the observer $R$, where $R$ has some effect on $S$ in virtue of  interacting with a system entangled with $S$, so a strict unitary description of $S$ relative to $R$ becomes inaccurate because it can't account for the effect produced by  $R$ itself. But since the entanglement is weak, the  effect on $S$ is also very weak, and hence the approximation where we describe $S$ relative to $R$ and leave $R$  out of our description is still \emph{mostly} accurate. Thus there occurs a partial collapse encoding the effect of the interaction with $R$ which cannot be modelled relative to $R$, but the disturbance is small and so the description in terms of unitary evolution is still fairly good. The fact that weak measurements demonstrate  exactly the  kind of partial collapse that we would expect given the picture in which quantum events arise from the breakdown of an approximation is surely an encouraging indication that the view adopted here is a sensible way of thinking about quantum events.

 \section{RQM \label{RQM}}

I will now consider how the ideas in the previous section might apply within the context of one specific relational approach: relational quantum mechanics (RQM). This  is a relational solution to the measurement problem originated by \cite{1996cr}, and it is distinguished in particular by its emphasis on avoiding anthropocentricism - a number of other relational or perspectival approaches  posit that quantum descriptions are  relativized to  conscious, and/or human, and/or macroscopic systems \citep{QBismintro, brukner2015quantum} but by contrast, RQM maintains that quantum descriptions can be relativized to any physical system.

To understand the motivation for such a view, it is helpful to consider how RQM resolves the Wigner's Friend scenario \citep{Wigner}. Suppose Wigner's Friend $F$ is measuring a quantum system $S$ in a coherent superposition of spin-Z states, $|\psi_S \rangle = \alpha | \uparrow \rangle_S +  \beta | \downarrow \rangle_S$. Presumably the Friend will see a definite outcome, but if Wigner watches from the outside and describes the whole interaction unitarily he will predict that after the measurement, $S$ and  Friend are in the following entangled state: 

\[ \psi_{SF} = \alpha |  \uparrow \rangle_S \otimes  | \uparrow \rangle_F   + \beta |  \downarrow \rangle_S  \otimes  | \downarrow \rangle_F  \]

Here, $| \uparrow \rangle_F $ is the state of the Friend corresponding to `having seen the $\uparrow$ result,' and likewise for $| \downarrow \rangle_F$. It appears that there is a contradiction between these two perspectives, since the Friend has seen a definite outcome, but to Wigner it appears as though the Friend is in a superposition of seeing two different outcomes. 

RQM addresses this paradox by saying that when the Friend measures $S$, a quantum event occurs in the quantum description relative to the Friend but \emph{not in the quantum description relative to Wigner} \citep{1996cr}. In this quantum event $S$ takes on a well-defined value of spin relative to the Friend and subsequently $S$ is in either the state $| \uparrow \rangle_S$ or $ | \downarrow \rangle_S$ relative to the Friend $F$, depending on which spin value occurred. But  because the quantum event is not part of the quantum description relative to Wigner, in that description unitary evolution continues, so the correct state relative to Wigner is $\psi_{SF}$ as shown above, and in principle Wigner could confirm this via tomographic measurements. Thus although the  perspectives of the Friend and Wigner are quite different in this scenario, they are both correct because a system can have different quantum states relative to different observers. All the versions of RQM that I will discuss in this article agree on this basic account of the Wigner's Friend scenario.

Note in particular that the reason this works is precisely because we have specified that the $S-F$ quantum event is part of the quantum description relative to $F$, but meanwhile  in the description relative to Wigner the evolution carries on unitarily as if no such event has occurred - that is, it is essential here that the $S-F$ quantum event does not play any role in the quantum description relative to reference frames other than $S$ and $F$. This basic motivation for the relational view will play an  important role in the discussion of section \ref{complete}.

\subsection{RRQM \label{RRQM}}

In this article, I will refer to the orthodox version of RQM  as RRQM\footnote{The RRQM/ARQM terminology comes from \cite{Faglia-2025}.}, and I will take it to be defined by the following postulates\footnote{The postulates of RQM have been stated in different ways in different presentations. The ones given here are not identical to any previous presentation but draw on those presented in \cite{pittphilsci19664} and \cite{Adlam-2023}; they are my best attempt at stating the central commitments of RRQM at the current stage in its development. }:

\begin{enumerate}

\item \textbf{Inherent relationalism:} All quantum descriptions are relativized to physical systems, and the correct quantum description of a system $S$ relative to a system $R$ is not always the same as the correct quantum description of $S$ relative to a different system $R'$.

\item \textbf{No Anthropocentricism:} All physical systems can play the role of `observers' to which quantum descriptions may be relativized. 

\item \textbf{Quantum events:} In any interaction between two systems $S, R$, a `quantum event' occurs in which the value of a variable of $S$ becomes definite relative to $R$. This value is selected from some basis with probabilities matching those assigned by the Born rule applied  to the quantum state of $S$ relative to $R$. The state of $S$  then undergoes an update in accordance with the usual quantum-mechanical state update rule. This is analogous to a `measurement' performed by $R$ on $S$, such that the value that becomes definite is the `outcome' of the measurement.   

\item \textbf{Completeness:} unitary quantum mechanics is complete and nothing needs to be added to the quantum formalism.
 
\item \textbf{Relativity of comparisons}: it is meaningless to compare a physical description relative to one system with a physical description relative to another system, except by invoking a third system relative to which the comparison is made.

\end{enumerate}

\subsection{ARQM and ARQM* \label{ARQM}}

In this article I will also discuss a recent variation on RQM known as ARQM.   As presented by \cite{Adlam-2023}, the purpose of this modification is to ensure that all observers share a common reality within which they can communicate and do science. It is is formulated by removing the \textbf{Relativity of comparisons} postulate  and replacing it with the following postulate:

\begin{enumerate}
 \setcounter{enumi}{4}
 
 \item  \textbf{Cross-Perspective Links (CPL):} In a scenario where some observer Alice measures a variable V of a system S, then provided that Alice does not undergo any interactions which destroy the information about V stored in Alice’s physical variables, if Bob subsequently measures the physical variable representing Alice’s information about the variable V, then Bob’s measurement result will match Alice’s measurement result.\footnote{This postulate follows the standard convention in RQM of using the word `observer' to refer generally to any physical system to which a quantum state can be relativized; it need not be assumed that Alice and/or Bob are conscious, human or macroscopic.}

\end{enumerate} 

However, I will now argue that as a matter of fact the  \textbf{CPL} postulate is probably inconsistent with the \textbf{Completeness} postulate. So in this article I will largely focus on a view that I will call ARQM*, which is similar to ARQM but which drops   \textbf{Completeness}\footnote{ It also permits some significant modifications to the \textbf{Quantum Events} postulate, but I will not count this as a modification, since the ongoing criticisms make it clear that the same is probably true for all formulations of RQM.}. I emphasize that at present ARQM* should not be addressed as a fully formulated interpretation but rather as a developing research program -  the current paper is intended as a contribution to that development, not as an unqualified defence of either ARQM or ARQM*. 

To see why we need to drop  \textbf{Completeness}, note that as pointed out by \cite{adlam2025kindrelationalitydoesquantum}, ARQM does seem to be committed to the existence of some facts about physical reality beyond what can be represented in the quantum formalism. For in virtue of the \textbf{CPL} postulate, ARQM is committed to the existence of some \emph{absolute} facts: once an $S-R$ quantum event has occurred, in principle all other observers can learn what happened during that event if they interact with $S$ or $R$ in the right way, so the facts about that event are absolute in the sense that they induce a set of modal facts which are the same for everyone. But ARQM nonetheless maintains the original commitment of RQM to the idea that quantum mechanics is inherently relational - it is \emph{only} suited to describe one system relative to another, so it cannot be used to characterize any absolute facts. It follows that ARQM is  committed to the existence of facts lying beyond what can be represented within the standard quantum formalism, i.e. some absolute facts. This suggests that we probably can't consistently maintain both \textbf{CPL} and \textbf{Completeness}, so if we want \textbf{CPL} or something similar to it we should move to ARQM* instead. 

We can see this in action in the Wigner's Friend scenario discussed above. Because of \textbf{CPL},  the $S-F$ quantum event is in some sense relevant to Wigner, because it determines what the outcome will be if Wigner performs a measurement $M$ aiming to establish the outcome of the Friend's measurement. However, ARQM still maintains that the \emph{quantum state} of $S$ and $F$ relative to Wigner at this time is $\psi_{SF}$ - that is, $\psi_{SF}$ correctly predicts the outcomes of all the other measurements that Wigner could perform on $S$ and $F$, and it also correctly predicts the statistics of the measurement $M$ over repeated experiments, even though a more accurate prediction can be given on any particular run of the experiment if we know about the quantum event $S-F$. We can see therefore that the facts about the $S-F$ quantum event and its relevance to Wigner are not represented in the standard \emph{quantum} description of $S$ and $F$ relative to Wigner, and therefore something beyond standard quantum mechanics will probably be required to properly characterize these features of ARQM.

Now, one might object to this argument on the grounds that we could find some kind of halfway house - i.e. perhaps we can separate out some part of the quantum formalism which is to be regarded as relational (e.g. the quantum states) while maintaining that some other part of the formalism is invariant across all relational descriptions and thus  can be used to define the absolute facts (e.g. the dynamics). But the problem with this approach is that in practice, quantum states and dynamics  are  intricately intertwined; for example, if an observer applies a certain Hamiltonian to a pair of qubits, the interaction between the qubits will depend strongly on the states of the environmental systems that the observer is using to apply the Hamiltonian, e.g. magnetic traps, sources of fields and so on. We could attempt to define some very abstract background Hamiltonian which is independent of the local states of specific systems, but it seems unlikely that such a thing would be adequate to derive quantum events, for as emphasized by  \cite{Faglia-2025}, a Hamiltonian on its own won't tell us which interactions take place and when they take place - this also depends on the states of the systems to which the Hamiltonian is applied. Additionally, \cite{Faglia-2025}  notes that we cannot possibly have a single global Hamiltonian in ARQM (or ARQM*) since the specification that all quantum states are relational entails that there is by definition no global quantum state on which such a thing could act. So we would instead have to work with  a collection of possibly overlapping Hamiltonians for different sets of systems, and in order to characterize structures common to all of them we would need some formal approach other than a Hamiltonian acting on a Hilbert space, so we would probably have to go outside of the quantum formalism. 

For these reasons it seems quite infeasible to cleanly separate out some aspect of the quantum formalism which can be regarded as `absolute' even when all of the quantum states are relativized, and which is nonetheless rich and  concrete enough to provide a precise characterization of the quantum events. Consequently it seems   most natural to say that the quantum formalism as a whole should be understood as a special kind of description which arises specifically in the context when we are characterizing one system relative to another, and  we require  some other kind of formalism to characterize the absolute facts. This makes sense because, as I will discuss further in section \ref{outside}, it is in any case to be expected that the process of scientific enquiry which has produced quantum mechanics will most immediately give rise to a theory which applies specifically to relative facts and which therefore may not be capable of describing absolute facts within its standard formalism. 

 Now, one might naturally worry that ARQM* is not really a version of RQM, since \textbf{Completeness} is a central feature of many presentations of orthodox RQM.  Ultimately I think this is not an important question: what matters is not whether ARQM* is `really' RQM but whether it is a plausible approach to the measurement problem in its own right. Still it should be noted that despite the differences mentioned above,   there are still significant commonalities between the two views - ARQM* maintains the central commitment to the idea that quantum mechanics is inherently relational, plus the idea that quantum events occur in interactions between any pair of systems, and as described above it also holds on to the original motivation for RQM in terms of the proposed relational resolution to the Wigner's Friend paradox.  Thus ARQM* shares important foundations with RRQM even if it also departs in significant ways from the original view. 

\subsection{Quantum events \label{partial}}

Much recent criticism of RQM has focused on the \textbf{Quantum events} postulate \citep{Mucino2022-MUCARQ,Calosi2024-CALRQM,Faglia-2025,Ladyman-2026}. The problem is that the postulate as stated above does not actually say how to derive a precise description of a quantum event from the quantum formalism, so a number  of questions still need to be answered. First,  it is not clear how to determine \emph{when} and \emph{under what circumstances} a quantum event takes place, and second, it is not clear how to determine \emph{in which particular basis} the value is selected in any given quantum event. 

As I will now explain, the proposal set out in section \ref{why} provides a natural answer to these questions. However it is not clear that this answer is available in the context of RRQM, due to the \textbf{Completeness} postulate. The proponents of RRQM usually understand \textbf{Completeness}  to mean that there are no physical facts of any kind beyond what can be captured in the quantum formalism, and hence RRQM cannot accept  that quantum mechanics is an approximation which works well as long as $S$ and $R$ are interacting only very weakly, since RRQM maintains there is nothing outside of quantum mechanics that it could be an approximation to. Thus RRQM presumably needs to be able to define the quantum events in a precise way entirely within the standard quantum formalism, because it does not admit any further descriptive resources.  For similar reasons, the proposal  is presumably not available within ARQM either. 

However, the proposal is quite natural within ARQM*, which as we have noted is in any case committed to the existence of absolute facts which lie outside of the quantum formalism. So in ARQM* we can simply posit that there exist absolute facts about reality which ground the effectiveness of relational quantum descriptions in certain circumstances, in much the same way as the complete observables of general relativity ground the meaningfulness of descriptions in terms of partial observables in certain circumstances  \citep{Rovelli_2002}. Indeed, within ARQM* the conclusion that a precise formulation of the quantum events must lie outside of the standard quantum formalism is in any case already obvious and natural - for as noted in section \ref{RQM}, it is essential to ARQM* that  quantum events are absolute, and since ARQM* is committed to the view that quantum mechanics is inherently relational, it follows that the absolute aspects of quantum events cannot be described using the quantum formalism, so we will inevitably need some other formalism to fully characterize them. 

Thus let us now see how this version of ARQM* can resolve questions about when and how quantum events occur. 

\paragraph{When does a quantum event occur? }

Clearly in this picture asking `when' a quantum event occurs is not a well-posed question. The relevant interactions  are always taking place - every system is continuously `measuring' every other system all the time. Asking `when does a quantum event occur?' really amounts to asking `when does the impact of the interaction become significant enough that the relational description is no longer accurate?' and it is evident that a question like that should not be expected to have a completely precise answer, since it is just a matter of whether the approximation is good enough for some purpose.

On this understanding of quantum events they are not generally pointlike: a quantum event continues for as long as $S$ and $R$ continue to interact strongly, which in principle could go on for an arbitrarily long period of time. However, many cases of interest to us involve a relatively short interaction. In particular, this is the case for many of the kinds of interactions that we would typically describe as measurements: the system $S$ and the measuring instrument $R$ start out interacting only very weakly,  and then  they are made to interact strongly for a short period of time for the purpose of performing a measurement, and then they go back to interacting weakly. In that case, within the description relative to $R$ we have a period of unitary evolution of $S$, plus a short period in which something happens to $S$ that cannot be described relative to $R$, and then we return to the unitary evolution of $S$. Because we had to leave the description relative to $R$, we get a non-linear `state update' or `collapse' within the description relative to $R$, and the period in which this takes place is our `quantum event.'  

At this point a natural question presents itself: if `quantum events' are not actually pointlike, why does it work so well to model wavefunction collapse as discrete and instantaneous? First of all, as originally argued by   \cite{von2018mathematical} there is typically a lot of latitude about where exactly in the chain of interactions reaching from the quantum system to the human brain we put in our wavefunction collapse -  \cite{Bell} called it the `shifty split' precisely because we can shift it around significantly without making any difference to the predictions. This suggests  that we can probably also stretch it out or compress it down to a point without making much difference to the predictions. In addition, in the picture suggested here a `quantum event' or `wavefunction collapse' is a representation of something that we simply can't model at all relative  to our current  reference system, so it  might as well be represented as instantaneous, since we can't say anything more concrete  about what the interaction actually involves in any case. So the fact that wavefunction collapse is often modelled as instantaneous does not imply that `quantum events' must literally be instantaneous.

\paragraph{Which interactions produce quantum events? }

Similarly, in this picture it is not really meaningful to ask  which specific  interactions give rise to quantum events: in some sense all of them do, but in many cases the effect of the interaction can be neglected or treated as a small perturbation. Asking `does a quantum event occur during this interaction?' really just amounts to asking `does this interaction ever have a significant enough impact that the relational quantum description becomes inaccurate?' 

 As noted in section \ref{nature}, we can distinguish a continuous spectrum of possibilities. If the interaction is very strong, the standard description in which $S$ evolves unitarily relative to $R$ becomes completely inapplicable - we cannot describe $S$ relative to $R$ in any way during this period, so we get a completely discontinuous nonlinear change, i.e. the kind of `collapse' appearing in the state update rule associated with a projective measurement. Whereas if the interaction is very weak, it can be disregarded completely, so the description of $S$ evolving unitarily relative to $R$ remains accurate. And in intermediate cases we will get some kind of partial collapse, as for example in the context of weak measurements.

\paragraph{In what basis is a value selected?}

 As noted above, in the picture I have advocated we can understand within the quantum formalism \emph{why} discontinuities should arise when a system interacts with a reference, but in order to explain precisely the nature of the resulting state transitions we would probably have to say more   about  the underlying absolute picture to which quantum mechanics is a relational approximation. 
Thus it is likely that a full understanding of the basis in which a value becomes definite during an quantum event probably requires going outside of the quantum formalism, and  therefore it is  not very surprising that attempts to identify the basis in question using only quantum states or dynamics run into problems \citep{https://doi.org/10.48550/arxiv.2107.03513,Mucino2022-MUCARQ}.

\paragraph{How do cross-perspective links form?}

The \textbf{CPL} postulate as stated above is formulated in quite an imprecise way - in particular, very little is said about the specific nature of the interaction required to implement a cross-perspective link. Additionally, in the account of ARQM* I have suggested here there is clearly a need to generalize the \textbf{CPL} postulate to deal with the fact that quantum events are not generally instantaneous - for example, to allow for partial collapses or scenarios in which the interaction between Bob  and Alice occurs while Alice and $S$ are still in the process of interacting.  
 
However, recall that the reason why the information encoded in a cross-perspective link is available across multiple perspectives is precisely because, according to ARQM*, it is absolute rather than relational. And I have emphasized that the absolute aspects of ARQM* cannot be characterized within the standard quantum formalism, which means that  the cross-perspective links belong to precisely the sector of ARQM* that we should not expect to be able to describe using only standard quantum mechanics. 

This is natural in any case due to the way in which quantum mechanics has been arrived at: when we do laboratory experiments, in the end all of our  measurements are performed relative to the same laboratory reference system and hence a theory formulated in this way will by default have little to say about the relationships between \emph{different} relative descriptions. So it is not very surprising that we would not be able to derive cross-perspective links in a way which is entirely internal to such a theory, and thus  it is likely inevitable that we will have to venture outside of standard quantum mechanics to describe these connections precisely.

\section{Can quantum events be derived internally to quantum mechanics? \label{complete}}

 As noted, the account of quantum events set out in the previous sections is presumably not available within RRQM due to the \textbf{Completeness} postulate. This leads to two difficulties for RRQM. 
 
 The first problem is that it would seem that RRQM simply has no explanation for the existence of quantum events at all:  they are simply postulated ad hoc as fundamental primitives, and they do not seem to arise in any natural way from the relational aspects of the picture, so there is an disunity to the approach which seems quite unsatisfying. As discussed in section \ref{why}, other kinds of relational view have access to a natural and satisfying explanation for `wavefunction collapse' or `quantum events,' but RRQM is not able to take advantage of such an explanation and this makes it seem like a  less appealing way of developing the relational approach.

The second problem is that the prospects of deriving the quantum events in a precise way entirely within the quantum formalism seem dim. This can be demonstrated by means of iterating examples, as in the work of  \cite{Faglia-2025} and \cite{Ladyman-2026} showing the failure of some natural proposals. But at a higher level we can also see directly from the basic principles of the relational approach  that it is quite unlikely that any such thing could work. The argument can be expressed as follows: 

\begin{itemize}

\item \textbf{Premise one:} Quantum mechanics is inherently relational, so if we are using the quantum formalism to derive an $S-R$ quantum event we must do so relative to some reference system. 

\item \textbf{Premise two:} Since systems are not assigned quantum states relative to themselves, we cannot describe an interaction between $S$ and $R$ or a joint state of $S$ and $R$ relative to either $S$ or $R$, and thus there is no possible way of giving a precise derivation of the $S-R$ quantum event relative to either $S$ or $R$.

\item \textbf{Premise three:} Since the $S-R$ quantum event does not appear in the quantum description relative to systems other than $S$ and $R$, it does not make sense to try to derive this event from the quantum description relative to some system other than $S$ or $R$. 

\item \textbf{Conclusion:} So the $S-R$ quantum event cannot be  derived in a precise way using the quantum formalism at all.

\end{itemize}

Below I will discuss some  nuances. However, I emphasize that my intention here is not to present a completely airtight argument - I refer to   \cite{Ladyman-2026} and \cite{Faglia-2025}  for detailed analyses of the  ways in which various specific attempts to derive the quantum events  fail. Rather my goal  is to offer some high-level motivation  illustrating that trying to derive quantum events entirely within the quantum formalism is simply not  very natural  given the commitments of the relational approach; comparatively, the picture in which quantum events correspond to the breakdown of the relational approximation seems significantly more coherent and natural.

\paragraph{Within RRQM}

In the context of RRQM specifically the argument is particularly simple. The $S-R$ quantum event is  supposed to occur relative to $R$\footnote{There is also an  $R-S$ quantum event relative to $S$, but here I will focus on the $S-R$ event.}. Because we can't describe $R$ relative to itself, we can't give  an explicit description of the interaction between $S$ and $R$ or assign a joint state of $S$ and $R$ within the description relative to $R$. Thus we cannot derive the $S-R$ quantum event entirely from the quantum description relative to $R$ because there is nothing that we could derive it from. 

Moreover, because RRQM is committed to \textbf{Relativity  of comparisons}, we also cannot use a description relative to some third-party external system $Z$ to derive the $S-R$ quantum event, because deriving a feature of the description relative to $R$ from a description relative to $Z$ would involve making a comparison between the descriptions relative to $R$ and $Z$. So it seems quite straightforward that, given the commitments of RRQM, there is no possible way to arrive at a precise derivation of a quantum event - except maybe  by appeal to a regress of relativization, which I will discuss shortly.

\paragraph{Beyond RRQM}

Even if we are willing to relax the \textbf{Relativity  of comparisons} postulate there are serious problems. We still can't derive the $S-R$ quantum event within a description relative to $S$ or $R$, and though in this case one could technically try  to derive the $S-R$ quantum event from a quantum description relative to some third party $Z$, it seems quite nonsensical to do so. For as  emphasized in section \ref{RRQM}, one of the central motivations for adopting a relational approach to quantum mechanics in the first place is precisely to be able to say that the quantum description of $S$ and $R$ relative to some third system $Z$ does \emph{not}  include the $S-R$ quantum event - this is essential to RQM's explanation of the Wigner's Friend scenario \citep{1996cr}. Thus, since the central idea of the relational construction is that the $S-R$ quantum event does not feature in the quantum description relative to $Z$, there is no reason to think that it can be derived in a precise way from this description. Indeed, one might worry that if it were so derivable then in a sense it \emph{would} be represented in the quantum description relative to $Z$, thus undermining the central motivation for the relational approach. So trying to derive the $S-R$ quantum event from a description relative to $Z$ seems like a very strange thing to do, given the motivating idea of the relational approach. 

  Moreover, if we nonetheless decide to attempt such a derivation, a further problem arises: the quantum description of $S$ and $R$  will be different relative to different external systems, and thus for any criterion we might apply to infer the presence of a quantum event, we will get different conclusions from different relative descriptions. Yet we are trying to determine whether a quantum event occurs within the single description relative to $R$ itself, and since most versions of the relational approach place emphasis on the equal validity of all relative descriptions, there does not seem to be any sensible way to choose some particular external frame to use for this purpose. So there seems to be no relationally acceptable way to derive a \emph{single} well-defined quantum event relative to $R$ in this context\footnote{\cite{Faglia-2025} considers the possibility of addressing this issue by using an existential quantifier - e.g. we could say that  a $S-R$ quantum event occurs relative to $R$ as long as we can derive such an event from at least one quantum description of $S$ and $R$ relative to an external system. But this criterion seems far too accommodating: unless we have some strong restrictions on how all the relative states are connected, it is plausible that it would almost always be met. Similarly one could imagine using a universal quantifier, but it is plausible that such a criterion would almost never be met.}.

\paragraph{A regress of relativization}

 In response to the two worries I have just raised, the proponents of RRQM might be tempted to appeal to a `regress of relativization' in the sense of \cite{Riedel_2024}. That is, they might  accept that it is impossible to derive  an $S-R$ quantum event relative to $R$ simpliciter, but nonetheless maintain that we can derive an $S-R$ quantum event such that the occurrence of the $S-R$ event relative to $R$ is itself secondarily relativized to the external system $Z$. Then, with regard to the worry about comparisons in RRQM, to derive an $S-R$ event relative to $R$ \emph{relative to $Z$} we would not need to compare the facts relative to $Z$ and $R$ in an absolute way but merely within the description relative to $Z$, which would not violate \textbf{Relativity of comparisons}. And with regard to the worry about which external reference system to use, we would simply have a different version of the $S-R$ quantum event for each external reference system, and hence no system would be privileged. 
 
 However, we are not going to be able to derive an $S-R$ quantum event relative to $R$ relative to $Z$ unless we first have a state of $S$ and $R$ relative to $Z$, and to obtain that we will need some quantum events involving $S$ and $R$ relative to $Z$, and presumably these will themselves have to be relativized to some other external system, and so on and so forth. Thus in this approach we will never reach a baseline at which we can stop and actually derive any  quantum events, so the regress picture presumably will not be appealing to someone who is concerned about being able to arrive at a  precise derivation of which quantum events actually occur.  
 
 Perhaps more importantly, this  seems like exactly the wrong approach given that, as noted above, the starting motivation for the relational picture was that an $S-R$ quantum event does \emph{not} exist within descriptions relative to external systems. The regress picture goes in precisely the opposite direction from this initial insight, because it tells us   that an $S-R$ quantum event cannot happen relative to $R$ or $S$ by themselves at all -  it can \emph{only} be derived relative to an external system, and that relativization must itself then be relativized to an infinite regress of external systems, so in some sense every $S-R$ quantum event eventually ends up relativized to everything \emph{except} the systems $S$ and $R$ themselves! Whether or not this is a viable view in its own right, it is certainly not a good realization of the original relational  idea that the $S-R$ event need not play a role in descriptions relative to systems other than $S$ and $R$.
 
 \paragraph{Some universal quantum feature}

 Given these difficulties, one might think that instead of deriving the $S-R$ quantum event within a  description relative to some particular reference system we should instead try to derive it from some  universal quantum structure which is independent of any particular choice of reference system. 

 However, this seems unworkable. For  in order to actually derive some quantum events we would need something reasonably concrete and non-trivial -  generic features of all quantum descriptions, such as `the evolution is unitary,' clearly will not be adequate. And yet in RRQM it would be impossible to find any more substantive universal feature more substantive, because in order to arrive at such a thing we would have to compare the descriptions across all reference frames in order to identify some shared structure, and such a comparison is forbidden in RRQM. That is, of course in RRQM we are not allowed to compare these descriptions in an absolute way, but in this case we cannot even compare them in a relativized way, since there is nothing outside of the set of  \emph{all} the possible reference systems to which this universal comparison could itself be relativized. So there is no possible way to arrive at such a universal feature even within some kind of relational description.
 
 Moreover,  even if we are willing to  relax \textbf{Relativity of Comparisons}, this approach probably still cannot work. For as explained in section \ref{ARQM}  it seems very unlikely that there is any way  to make a clean split in such a way as to identify some universally valid aspect of the quantum formalism which is the same across all relative descriptions and yet rich enough to ground the occurrence of quantum events, and thus it is implausible that there exists any  universal quantum structure which could be used to derive the quantum events.

\subsection{Responses}

From the argument above it is evident at a heuristic level that trying to derive quantum events entirely within the quantum formalism is not a very sensible thing to do, given the basic commitment to the idea that quantum mechanics is inherently relational. This high-level insight is supported by detailed investigations carried out by  \cite{Faglia-2025} and \cite{Ladyman-2026} demonstrating that attempts to perform such a derivation encounter numerous problems, as we might expect in light of the argument above.

 So if the proponent of RRQM wishes to resist departing from the quantum formalism, what can they do? Perhaps the most obvious possibility is to deny the claim that  systems cannot be assigned quantum states relative to themselves. In that case we can in fact describe an interaction involving $S$ and $R$ within a description relative to $R$, and then perhaps it would be possible to derive an $S-R$ quantum event from within the quantum description relative to $R$ alone. However, we saw in section \ref{why} that there are a number of substantive  reasons to think this should not be done in relational approaches in general. Moreover,  \cite{Faglia-2025} offers a further reason why it should not  be done in RQM specifically: one natural reading of RQM  suggests that a system $S$ acquires a quantum state relative to $R$ only as a result of  an $S-R$ quantum event, and thus since the quantum events always involve two distinct entities, it simply is not  possible for a system to have a state relative to itself.

 An alternative possibility would be to accept the argument  above and maintain that this means  quantum events  are not defined precisely at all - there must be a kind of real, fundamental nomic vagueness which inheres in the world itself as a direct consequence of  the metaphysical commitments of orthodox RQM - see \cite{Chen_2022} for some discussion of the concept of fundamental nomic vagueness. For example, perhaps a proponent of a version of RRQM which involves a regress of relativization  in the sense of \cite{Riedel_2024} might be drawn to  such a picture.
 
 But this is not a very appealing response, because as we have previously noted, within a relational approach which insists on \textbf{Completeness} the existence of the quantum events itself does not seem to follow in a particularly natural way from the underlying metaphysical commitments - as noted, they seem to be simply postulated ad hoc to save the phenomena. And if the quantum events are both an ad hoc addition to save the phenomena and also they cannot even be defined in a clear and precise way, the overall picture looks quite unappealing. In addition, the postulation of fundamental nomic vagueness should always be approached with care, for if applied in too profligate a way it might encourage us to  leave things vague where in fact attempting precision could give rise to  valuable new progress.  

A final possibility is to accept the argument as written and maintain that RQM does not actually need `quantum events.' This is the route taken by \cite{dibiagio2026relativeinformationrelativefacts}, in a version of RQM which  explicitly gives up on the possibility of representing or modelling quantum events either absolutely or relationally. Rather we simply say that a `measurement' of $A$ by $B$ is simply the continuous unitary evolution of $A$ and $B$ together as described relative to some third party system $C$: `the information about the measured variable relative to the apparatus grows continuously ... there is no special moment in which the information that the apparatus has about the system jumps from zero to a finite value.'  

The problem with this response is that the whole point of the measurement problem is that we need to understand how quantum measurements can give rise to the appearance of single definite outcomes; in a relational approach  of course we may wish to say that  these single outcomes exist only relative to specific observers, but nonetheless we must end up with individual observers perceiving single outcomes or else the measurement problem has not been solved.  And the approach set out by  \cite{dibiagio2026relativeinformationrelativefacts} simply does not yield anything like a single definite outcome: from the point of view of $C$, the systems $A$ and $B$ simply end up in a superposition of seeing all the outcomes together. Of course,  $C$ can then perform  a measurement on $A$ and/or $B$, e.g. by asking $A$ what the outcome was, but that is no help because the formalism gives us no way to model the occurrence of a single outcome in that measurement either, except by moving to a description relative to some fourth system $D$, which of course will simply reproduce the same problem. At no point in this picture do we ever arrive at a single definite outcome of any measurement, \emph{even relative to an individual observer.} 

One might perhaps try to argue that, in some sense which is difficult to articulate, the existence of a correlation between $A$ and $B$ relative to $C$ `corresponds,' from the point of view of $A$, to a definite measurement outcome relative to $A$. But this cannot be right, since for a given measurement interaction the correlation is always the same whereas for most measurements the \emph{outcome} is not always the same: we have to add something over and above the correlation to make sense of the stochastic selection of one actual outcome.  

 \cite{dibiagio2026relativeinformationrelativefacts} try to make the case that the scenario they describe is just the same as classical measurement because `in classical information theory, measurements of a discrete variable exhibit the same apparent tension between discrete outcomes and continuous information acquisition: while the measurement result needs to be one of a finite set of values, the relative information between the measured variable and the register grows continuously in time.' But this is a misdiagnosis. Of course it is true that in a classical measurement information is acquired in a continuous way, but that is because classical measurements just reveal the preexisting value of a variable which has had that value all along, so what is continuously acquired in such a case is simply information about that value. Whereas in RQM, since RQM does not permit hidden variables there is usually no preexisting value of the relevant variable and so we cannot make do with a continuous growth of correlation - if the measurement has only one outcome (relative to an individual observer) at some point in the process of measurement there needs to be a stochastic selection of one of those outcomes (relative to an individual observer).  The relevant tension, therefore, is not between `discrete outcomes and continuous information acquisition,' but between \emph{single} outcomes and the existence of multiple possibilities up until the time of the measurement, so actually this scenario is quite disanalogous to a conventional classical measurement.

Perhaps \cite{dibiagio2026relativeinformationrelativefacts} ultimately intend to give up on the occurrence of single definite outcomes and to say that even relative to a given observer all the outcomes of the measurement always occur. In that case, they are saying that RQM is just a version of the Everett interpretation \citep{Everett} - after  all, the characterization of measurement as nothing more than a process in which instrument and system become correlated beyond a certain degree is more or less exactly how Everettians describe the process of quantum measurement \citep{Everett,Wallacebook}. There remain some differences at the level of interpretational gloss, but it is unclear how much those differences matter; certainly at this point the claim of RQM to offer a substantive alternative to the many-worlds picture seems significantly undermined. In particular, a central difficulty for the Everett interpretation involves the problem of accounting for probabilities \citep{price2008decisions,AdlamEverett}, and since this difficulty arises precisely as a result of the fact that Everettian measurements do not have single definite outcomes, the version of RQM described by \cite{dibiagio2026relativeinformationrelativefacts} will face exactly the same problems, so if this is the route that proponents of RQM want to take then there is an urgent mandate for them to explain how they plan to solve the analogue of the Everettian probability problem.

\subsection{Going outside the formalism \label{outside}}

One response I foresee to this argument is that if it's really true that a satisfactory relational approach will inevitably lead us outside of the quantum formalism,   that means relational approaches require a great deal of new technical work and potentially new physics, and this is a prima facie reason to reject them altogether. 

 I do not find this response convincing, because it is in any case clear that there exists physics of some kind beyond quantum mechanics, so there is no  reason to think that the correct interpretation of quantum mechanics will be definable in a completely precise way without any appeal to physics beyond quantum mechanics. If quantum mechanics itself is just an approximation to some other physics, as we have very good reason to think it must be, then it makes little sense to insist that an acceptable interpretation of the theory can be stated in a completely precise way without appeal to other physics. Of course one should not unreservedly accept an interpretation which relies on as yet undiscovered physics, but as I see it the question at issue is not about whether relational approaches should be \emph{accepted} right now but rather whether some version of the relational programme can inform a viable and interesting ongoing research program, and I think there are good reasons to give an affirmative answer to this question.

 Of course, that does not mean that we should find it reasonable for a putative interpretation of quantum mechanics to simply gesture vaguely at `unknown physics' to plug holes in the approach. But the picture I have set out here does not do this: rather it explains clearly  why the relational picture must have  `quantum events' and why it is inevitable that their complete description takes us outside of the quantum formalism, including concrete indications of the direction which the research program should be taken. 

 In particular, one reason to think this is an interesting direction to pursue is  that it may be  be connected  with ongoing research on general relativity and quantum gravity. I noted in section \ref{why} that the story in which  quantum events are consequences of the breakdown of an approximation due to the strength of the system-reference interaction is in fact quite similar to a scenario that has already been studied within general relativity. Moreover, in the context of general relativity, quantum gravity and the quantum reference frame formalism, physicists already have a number of well-developed techniques for moving between absolute and relational descriptions \citep{Rovelli_2002,Miyadera_2016,Loveridge_2018,delahamette2021perspectiveneutral},  so there are  plenty of good indications for how we might proceed in order to develop the kind of formalism needed. 

 Furthermore, although the approach explored here does require that we  give up \textbf{Completeness}, it is arguably more appealing than standard  hidden variable approaches. This is because  in this picture the explanation for why something important is missing from quantum mechanics does not rely on stipulations that simply hide some part of nature from us in an ad hoc or conspiratorial way - rather it is a   natural  consequence of the very nature of scientific enquiry that some absolute facts should have been left out of quantum mechanics proper. 
 
 This is because we necessarily study physics relationally - systems are measured relative to our measuring devices, localized relative to us, evolving relative to our clocks - and thus it is to be expected that a theory such as quantum mechanics, which was first and foremost designed to accommodate experimental observations, will end up  functioning primarily as a characterization  of \emph{relative} facts and may therefore leave out some absolute facts. In particular, as noted in section \ref{partial}, it is natural that a theory of this kind will omit absolute facts about the connections between different relative descriptions, i.e. precisely the kinds of facts which I have previously suggested must be sought outside of the quantum formalism. Therefore the postulation of some physics beyond standard quantum mechanics to which the relative descriptions are an approximation seems quite well-motivated  in this picture.

\section{Specific Criticisms \label{criticisms}}

I will close by commenting more specifically on two recent works which have raised worries about quantum events in RQM. Both   \cite{Ladyman-2026} and \cite{Faglia-2025} make similar assumptions and hence receive roughly the same response. That is, both works assume that quantum events are discrete and instantaneous and involve total collapses, and both assume that a valid formulation of RQM must derive quantum events entirely from within the standard quantum formalism. These assumptions are  reasonable, since they reflect things that proponents of RQM have actually said. However, we have seen here that careful reflection on the nature of `quantum events' should prompt us to give them up in any case, and thus at least in the context of ARQM* objections based on these assumptions do not  have purchase.  

I will comment here on a few points from these papers which  seem potentially relevant to  the account of quantum events that I have proposed.

\subsection{Ladyman and Thompson}

 First of all, note that   \cite{Ladyman-2026} do not simply take for granted that quantum events are instantaneous; rather they argue that this is essential because  RQM's solution to the measurement problem hinges on the idea that measurements are quantum events, so `the discrete and instantaneous nature of measurements in standard quantum theory automatically means that quantum events must also be discrete and instantaneous.' One might worry that this argument will also cause problems for the proposal here, since I have suggested that quantum events are not discrete or instantaneous. 
 
 But in fact, measurements in `standard quantum theory' are not really discrete and instantaneous either:  real measurements involve continuous physical couplings between the measuring device and the system of interest which can be prolonged for an indefinitely long period of time \citep{jordan2024quantum, Katz_2008}. So in fact, a picture which associates quantum events with continuous physical couplings is a more natural fit to modern measurement theory in any case, as we saw in section \ref{partial} particularly with respect to the phenomenon of partial collapse.  It is true that the `wavefunction collapse' part of the measurement process is typically modelled as discrete and instantaneous, but as explained in section \ref{partial}, this does not imply that the quantum events must be instantaneous. 

 Next, Ladyman and Thompson discuss a suggestion made by \cite{Adlam-2023} to the effect that a precisification of the notion of when quantum events occur might require appeal to some features of quantum gravity, and criticize this suggestion on the grounds that `it seems we should view RQM as at least as speculative as quantum gravity. In that case RQM cannot be claimed to be a solution to the measurement problem in the absence of a working theory of quantum gravity.'  Presumably then they would respond in the same way to the suggestion here that a full account of quantum events requires passing beyond the quantum formalism. 

However, as discussed in section \ref{outside}, I think it is a mistake to dismiss any interpretation of quantum mechanics which  makes reference to possible future developments in physics in this way. While we should be suspicious of vague and unmotivated appeals to future physics, in this case there are clear reasons why such a thing is needed and also a number of existing technical formalisms which offer a starting point in thinking about  how to actually develop such a thing. Under such circumstances the optimal approach is  to allow the problem of quantum interpretation and the search for new physics to interact and pass ideas back and forth between them, rather than insisting on a strict separation.

 Finally,  \cite{Ladyman-2026} consider an observer $O$ witnessing an interaction between two qubits $q_1$ and $q_2$, and maintain that resolving the problem of quantum events requires answering three questions. So I will now say what the answers to these questions would be, if we accept the approach I have advocated in this article:

\begin{itemize}

\item \textbf{ From the perspective of observer $O$, when does the quantum event of `$q_1$ and $q_2$  interacting' occur? }

RRQM, AQRM and ARQM* all specify that this quantum event is not represented in the \emph{quantum} description relative to $O$: in that description the quantum state of the two qubits $q_1$ and $q_2$ remains uncollapsed and contains no reference to any specific outcome of the quantum event. And since RRQM does not posit  cross-perspective links nor any facts beyond relational quantum descriptions it need not say anything further on this matter. 

 Here there is some ambiguity in the langauge used by Ladyman and Thompson. They write, `Since a modal fact becomes true for $q_2$ at the moment that the interaction happens relative to $q_2$, it would be natural for ARQM to identify the timing of the quantum event `$q_1$ and $q_2$  interacting'  relative to $O$ with the moment that the same interaction takes place relative to $q_2$.'  But this is not quite  correct, because in fact this quantum event will  \emph{never} be represented within the quantum description of $q_1$ and $q_2$ relative to $O$, not even if we are using ARQM or ARQM*. This is because, as I have emphasized already, a central motivation for RQM is precisely the idea that the $S-R$ event does \emph{not} exist within the quantum description of $S$ and $R$ relative to external systems. Perhaps Ladyman and Thompson have in mind some kind of hybrid description relative to $O$ which includes both the relative facts encoded by the quantum formalism and also some absolute facts about the quantum events, but it is not at all clear what such a thing would look like (and in that case it is a little misleading to ask about when the event occurs `relative to $O$,' because  an absolute event does not take place `relative to' anything, except in the trivial sense in which an absolute fact could be described as being   relative to everything).
 
 However, it is of course true that  ARQM*  will have to say something about when the \textbf{CPL} postulate takes effect, i.e. after which time will an appropriate interaction between $O$ and $q_2$ yield information about the value of $q_1$ that became definite relative to $q_2$ in the interaction between $q_1$ and $q_2$? Roughly speaking, the answer is presumably that  the cross-perspective link will simply yield whatever information about $q_1$ is available to $q_2$ at the time of the interaction between $O$ and $q_2$, which  may be partial and incomplete if the interaction is weak or ongoing. As noted in section \ref{partial}, a more precise formulation of the \textbf{CPL} postulate would probably require going outside of quantum mechanics, and thus it is not possible to give a more concrete answer to this question using only the quantum  formalism.

\item \textbf{From the perspective of system $q_2$, when does the quantum event of `$q_1$ and $q_2$  interacting' occur? }

In a sense, the answer is `always' -  the qubits will effectively always be undergoing some kind of interaction and so there is always some small effect of $q_1$ on $q_2$ which disrupts the unitary evolution. However, much of the time these effects can be neglected. So to say under what circumstances something that we would call a `quantum event' occurs, we simply need to decide when the interaction between $q_1$ and $q_2$ has become so strong that it no longer works well to describe $q_1$ on its own, independent of $q_2$. Clearly there will not be any  precise answer to this question - the approximation does not fail all at once, it simply becomes worse as the interaction becomes stronger, so there is no exact moment at which the `quantum event' begins. 

Ladyman and Thompson do consider the hypothesis that quantum events take place whenever an interaction occurs, and conclude that it can't work because `quantum events lead to the collapse of the wave function for the quantum state relative to the observers involved in the interaction,' so this  `would make it impossible for quantum systems to remain coherent as long as they have any interaction whatsoever with the external environment, which is implausible and contrary to experimental practice.' 

Now first of all, this formulation of the argument does not seem right, because what RQM tells us is that when a quantum system $S$ interacts with an environment $E$, we get a quantum event involving $S$ and $E$ and thus $S$ will lose some coherence \emph{relative to $E$}. Yet what is seen in `experimental practice' corresponds to the observations of a third party observer $O$ watching the system interacting with the environment, so what is actually observed is whether or not the system loses coherence \emph{relative to $O$}. And as noted above, the motivating idea of of RQM is precisely that when $S$ loses coherence relative to $E$ it does not necessarily lose any coherence relative to a third party $O$ - that is how RQM resolves the Wigner's Friend paradox. So whether or not $S$ loses coherence relative to $E$ is not relevant in any direct way to the observations made in experimental practice, at least until we add further specifications connecting the relational descriptions relative to $O$ and $E$.

 But in any case, the worry raised here is  predicated on the picture of quantum events as discrete instantaneous events involving a total collapse of the wavefunction. The same issue does not arise in the picture suggested in this article where a `quantum event' is merely a way of describing  a scenario in which  a certain  approximation becomes inaccurate -  the fact that every interaction renders the approximation inaccurate to some degree need  not lead to widespread loss of coherence, because in many cases the disturbance from the interaction is sufficiently small that the approximation remains passably good and so the coherence remains.

\item \textbf{What is  the relationship (if any) between the quantum events that occur relative to the observer $q_2$ and relative to the observer $O$?}

RRQM   says that there is no such relationship. ARQM* uses the \textbf{CPL} postulate to give a high-level heuristic which connects quantum events involving $q_1$ and $q_2$ with quantum events  involving $q_2$ and $O$. At this stage it is probably not possible to give more than a high-level heuristic, because as noted in section \ref{partial},  it is unlikely that a fully  precise version of \textbf{CPL} can be formulated in a way that is   internal to the quantum formalism.

\end{itemize}

\subsection{Faglia}

 As  with Ladyman and Thompson, \cite{Faglia-2025}  assumes that quantum events must be defined from within the quantum formalism, then considers a range of possibilities and demonstrates that none of them succeeds. Given the argument of section \ref{complete}, this is to be expected. Indeed I take the work of \cite{Faglia-2025} as a useful contribution toward demonstrating that indeed, the absolute facts required for the formulation of ARQM or ARQM* cannot  be described entirely in the quantum formalism, as discussed in section \ref{ARQM}.

Note that Faglia does consider the possibility of seeing quantum events as approximations; however, most of this discussion  does not really apply to the proposal here. For example, Faglia writes `The occurrence of a (quantum event) cannot depend on a possibly arbitrary decision on what level of approximation is appropriate, if (quantum events) are to play such a fundamental role in the theory.'\footnote{In this quote I have replaced the word `interaction' with `quantum event' because as discussed throughout this article,  interactions are ubiquitous and continuous and we probably only want to use the term `quantum event' to refer to the case where the interaction is sufficiently strong.} 
This seems very reasonable if `quantum events' are considered as the fundamental ontology of the theory. But as Faglia goes on to point out, the presence of approximations is not necessarily a problem if the approximation is understood as simply an epistemic feature of our imperfect attempts to determine which quantum events occur. In particular, if we understand the quantum formalism itself as a relational approximation to an underlying absolute description, as suggested in this article, then it is very natural that the definition of the quantum events should also be approximate.

Additionally, Faglia writes that `if RQM’s interactions need to be identified via approximate conditions, then they do not naturally arise from the formalism of quantum theory. The supporters of RQM thus need to show that the postulation of events is not ad hoc.' However, as we saw in section \ref{why} the occurrence of discontinuous quantum events actually \emph{does} arise in a very natural way from the formalism of quantum theory, understood  relationally. That is, although I agree that we would need to go outside the formalism of quantum theory to fully \emph{characterize} these events in a precise way, the fact that such events occur is fully explainable in a way that is internal to quantum theory interpreted relationally, so it is not ad hoc at all.

 \section{Conclusion}

This article began with the observation that  within a relational picture there is an obvious reason why the unitary quantum description should be expected to break down when a system $S$ interacts with the system $R$ relative to which we are describing it, thus explaining why  `wavefunction collapses' or `quantum events' are inevitable in a relational account of quantum mechanics. The ability to give a clear explanation for wavefunction collapse and a straightforward physical characterization of what is so special about `measurement' is actually a very appealing feature of the relational view. 

We then saw that once we understand why quantum events occur, we can easily answer questions about the correct definition of quantum events. This approach suggests that quantum mechanics must be understood as an approximate relational description which arises in the case where the system and reference are interacting only weakly, so a `quantum event' simply refers to the circumstances in which that approximation breaks down. As with all approximations there is no exact, well-defined line between circumstances where the approximation is accurate and circumstances where it is not, so it is a mistake to seek an exact formulation of when  and under what circumstances a `quantum event' occurs.  

However, this line of argument suggests that the critics of  RQM are correct in the following sense:  it is probably true that we will never be able to define quantum events in a fully satisfactory way using only the quantum formalism and nothing else. Thus these worries offer interesting clues about how the relational approach ought to be further developed - i.e. toward the picture in which quantum descriptions are relational approximations to an underlying absolute reality, perhaps characterized by something like the `complete observables' featuring in existing formulations of general relativity. I hope that recent work demonstrating explicitly the failure of attempts to derive quantum events from the quantum formalism will serve as an impetus for development in this direction. 

\section{Acknowledgements}

 This work was supported by the  John Templeton Foundation Grant ID 63209.

\end{document}